# Abelian Chern-Simons gauge theory on the lattice

Bingnan Zhang[*]

*Department of Physics and NHETC, Rutgers University, Piscataway, New Jersey 08854, USA*

The Abelian Chern-Simons gauge theory is constructed on the three-dimensional spacetime lattice. This proposal introduces both lattice and dual lattice, and the gauge field on the dual lattice is expressed in terms of the gauge field on the original lattice. This treatment circumvents the issue of forward/backward difference, which is the common problem that many previous proposals have, and also avoids the duplication problem, which prevents people from introducing the dual lattice. The form of the lattice action is very simple and is symmetric with respect to the three spacetime dimensions. These features make it straightforward to calculate the expectation values of Wilson loops, and the results agree with the topological field theory in continuous spacetime. Generalizations to multiple types of lattices are also discussed.

## I. INTRODUCTION

As the simplest topological field theory, Chern-Simons theory [1] with Abelian gauge group has a clear geometrical meaning: it counts the linking number of loops [2,3]. However, there are ambiguities regarding the self-linking, and singularities also arise when loops intersect [2,4]. The former is called "framing anomaly", and it may have important contributions to thermal Hall conductivity, Hall viscosity, and linear responses of fractional quantum Hall systems [5,6]. One way to regularize such ambiguities is to define the theory on a lattice. The lattice approach also makes it possible to study Chern-Simons theory with discrete gauge groups like $\mathbb{Z}_n$, which has no clear analog in continuous spacetime. Moreover, the lattice construction of Chern-Simons theory can also be used to formulate a composite fermion description for fractional Chern insulators [7].

Lattice Chern-Simons theory has been studied for decades [8–19]. Gauge invariance and duplication are the two major problems that prevent us from having a satisfactory lattice implementation of Chern-Simons theory with Abelian gauge group. For proposals that do not introduce the dual lattice, the straightforward idea of replacing differentiation with lattice difference fails, because forward difference turns into backward difference under summation by parts, and the theory is no longer gauge invariant. One either has to invent very complicated versions of lattice difference [12,13] or introduce extra terms in the action to make it gauge invariant [16]. Introducing the dual lattice solves the forward/backward difference problem, and the theory is gauge invariant. However, extra gauge fields and matter fields have to be introduced. The theory reduces to *adb* Chern-Simons theory rather than *ada* theory in the small lattice spacing limit [9]. In Ref. [15], a straightforward construction is provided using the algebraic geometry language. However, this method only applies to simplex lattice, and it cannot be generalized. Moreover, the lattice action only agrees with the Abelian Chern-Simons action in continuous spacetime at leading order in the small lattice spacing limit. In Ref. [8], a very general construction of Abelian Chern-Simons theory on any proper spatial lattices is proposed. However, the time direction has to be continuous, and the action is not symmetric with respect to the three dimensions, which makes the calculation of Wilson loops that involve time components complicated. In this paper, a new definition of lattice Abelian Chern-Simons theory is proposed. This proposal introduces both lattice and dual lattice, and the gauge field on the dual lattice is expressed in terms of the gauge field on the original lattice. This treatment circumvents the issue of forward/backward difference and also avoids the duplication problem, which prevents people from introducing the dual lattice. The matter field can be regarded as a product of the parton fields on the vertexes of a lattice cell, and the Wilson loop consists of multiple lines on the lattice. This regularizes the divergence in the continuous theory. The form of the action is symmetric with respect to the three spacetime dimensions, and the calculation of Wilson loops is very straightforward. This construction works for both compact and noncompact gauge groups.

This paper consists of several parts. Section II presents the basic definitions of Abelian Chern-Simons theory on cubic lattice. Section III studies the change of the action

---

under gauge transformation. It is shown that lattice difference preserves its form under summation by parts. Section IV calculates the expectation values of Wilson loops using the saddle point method and shows that the results agree with the continuous theory. Section V restores the lattice spacing $\epsilon$ and studies the error between the lattice theory and the continuous theory as we take the small $\epsilon$ limit. Section VI goes to the Hamiltonian formalism and counts the dimension of the Hilbert space. Section VII generalizes the construction to triangular prism lattice and hexagonal prism lattice. Section VIII is a summary.

## II. THE DEFINITION

In continuous three-dimensional spacetime, the Chern-Simons Lagrangian is

$$\mathcal{L}_{CS} = \frac{k}{4\pi}\epsilon^{\mu\nu\rho}a_\mu\partial_\nu a_\rho. \qquad (1)$$

Once the lattice is introduced, $a_\mu$ becomes a link variable, and $\epsilon^{\mu\nu\rho}\partial_\nu a_\rho$ becomes the flux that penetrates a plaquette. In this paper, Abelian lattice Chern-Simons theory is first defined on the cubic lattice, and then, the construction is generalized to polygonal prism lattices. The lattice action is

$$S_{CS} = \frac{k}{4\pi}\sum_{\text{plaquette }p}\bar{a}_p\phi_p, \qquad (2)$$

where $\bar{a}_p$ lives on the dual link that penetrates plaquette $p$, which is shown in Fig. 1. This is different from Kantor and Susskind's construction in [9], because here $\bar{a}_p$ is not another gauge field but the average of the $a$ field on the eight vertical links surrounding plaquette $p$. In Fig. 1, the $\bar{a}$ field penetrating the plaquette $ABCD$ is

$$\bar{a}_{ABCD} = \frac{1}{8}(a_{AE} + a_{BF} + a_{CG} + a_{DH} + a_{IA} + a_{JB} + a_{KC} + a_{LD}), \qquad (3)$$

and the flux is

$$\phi_{ABCD} = a_{AB} + a_{BC} + a_{CD} + a_{DA} \qquad (4)$$

through the plaquette $ABCD$.

Matter field lives on the dual lattice, and it couples to the gauge field on the dual link (Fig. 1),

$$S_{\text{couple}} = \sum_{\text{adjacent cubes}}\bar{\psi}_{II}e^{i\bar{a}}\psi_I + \text{H.c.} \qquad (5)$$

Here, we have set the charge to 1, and the gauge group is compact. If the charge $Q$ is not 1, we just insert a $Q$ in front of $\bar{a}$ in Eq. (5). If we allow $Q$ to take continuous values,

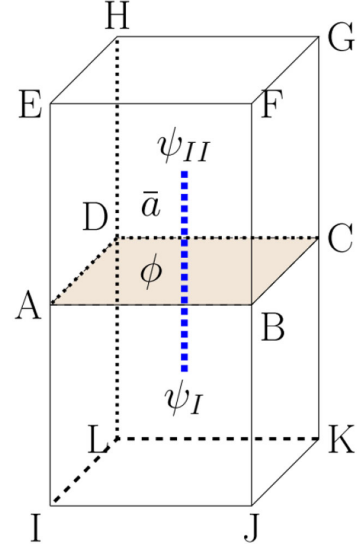

FIG. 1. The $\bar{a}$ field penetrating a plaquette is the average of the $a$ field on the eight vertical links surrounding the plaquette. Matter field lives on the dual lattice, and it couples to the gauge field on the dual link.

then the gauge group is noncompact. However, this does not affect the construction.

## III. GAUGE INVARIANCE

Gauge invariance is one of the key standards that measure whether a lattice construction of Chern-Simons theory is correct or not. This section checks the invariance of the lattice action defined in Eqs. (2) and (5) under small and large gauge transformations. The plaquettes and links have three independent directions, and every $a$ field on a link couples to eight plaquettes. These eight plaquettes are the plaquettes connected with and perpendicular to the link. This can be seen by substituting Eq. (3) into Eq. (2). In order to check the gauge invariance, we first pick out one direction and suppose every $a$ field on a link only couples to two plaquettes. This is shown in Fig. 2. Under the lattice gauge transformation

$$a_i \to a_i + \Delta\lambda_i = a_i + \lambda_{i+1} - \lambda_i, \qquad (6)$$

the change of the sum

$$S_1 = \sum_{i=1}^{n}(a_i + a_{i-1})\phi_{p_i} \qquad (7)$$

is

$$S_1 \to S_1 + \sum_{i=1}^{n}(\lambda_{i+1} - \lambda_{i-1})\phi_{p_i} = S_1 - \sum_{i=1}^{n}\lambda_i(\phi_{i+1} - \phi_{i-1}), \qquad (8)$$

where a periodic or vanishing boundary condition has been assumed. Define

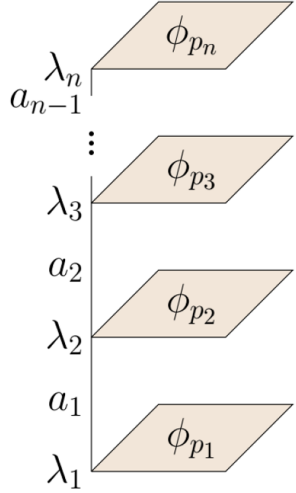

FIG. 2. Pick out one direction and suppose every $a$ field on a link only couples to two plaquettes.

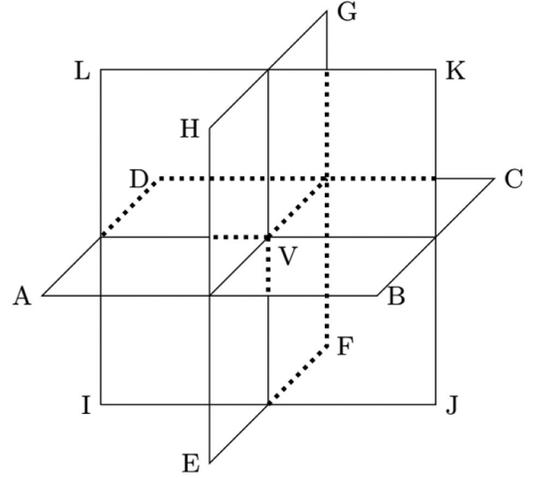

FIG. 3. On the cubic lattice, every vertex (vertex V in this figure) couples to 12 plaquettes.

$$\Delta^* f_i = f_{i+1} - f_{i-1}, \quad (9)$$

and Eq. (8) tells us

$$\sum_{i=1}^{n} \Delta^* \lambda_i \phi_{p_i} = -\sum_{i=1}^{n} \lambda_i \Delta^* \phi_{p_i}. \quad (10)$$

So the change of $S_1$ can be written as

$$\delta S_1 = -\sum_i \lambda_i \Delta^* \phi_{p_i}. \quad (11)$$

Unlike the literature [12], summation by parts is allowed in this proposal, and there is no forward/backward difference problem.

Sum Eq. (11) over the three directions, and letting every link couple to eight plaquettes, we have

$$\delta S_{\text{CS}} \propto -\sum_V \lambda_V (\Delta^* \phi_{ABCD} + \Delta^* \phi_{EFGH} + \Delta^* \phi_{IJKL}), \quad (12)$$

where the labels are specified in Fig. 3, and $\Delta^*$ is defined in Eq. (9).

As indicated in Fig. 4,

$$\Delta^* \phi_{ABCD} + \Delta^* \phi_{EFGH} + \Delta^* \phi_{IJKL}$$
$$= \phi_{ONRS} - \phi_{PMQT} + \phi_{MQRN}$$
$$- \phi_{PTSO} + \phi_{PMNO} - \phi_{TQRS}$$
$$= 0 \quad (13)$$

because they form a closed surface, and every edge is canceled. So $S_{\text{CS}}$ is gauge invariant.

The matter field at the center of a cube can be regarded as a product of the parton fields on the cube's eight vertexes. In Fig. 1, the matter field gauge transforms as

$$\psi_I \to \psi_I e^{\frac{i}{8}(\lambda_A + \lambda_B + \lambda_C + \lambda_D + \lambda_I + \lambda_J + \lambda_K + \lambda_L)},$$
$$\bar{\psi}_{II} \to \bar{\psi}_{II} e^{\frac{-i}{8}(\lambda_E + \lambda_F + \lambda_G + \lambda_H + \lambda_A + \lambda_B + \lambda_C + \lambda_D)}, \quad (14)$$

while $\bar{a}_{ABCD}$ defined in Eq. (3) transforms as

$$\bar{a}_{ABCD} \to \bar{a}_{ABCD} + \frac{1}{8}(\lambda_E + \lambda_F + \lambda_G + \lambda_H + \lambda_A + \lambda_B$$
$$+ \lambda_C + \lambda_D) - \frac{1}{8}(\lambda_A + \lambda_B + \lambda_C + \lambda_D + \lambda_I$$
$$+ \lambda_J + \lambda_K + \lambda_L). \quad (15)$$

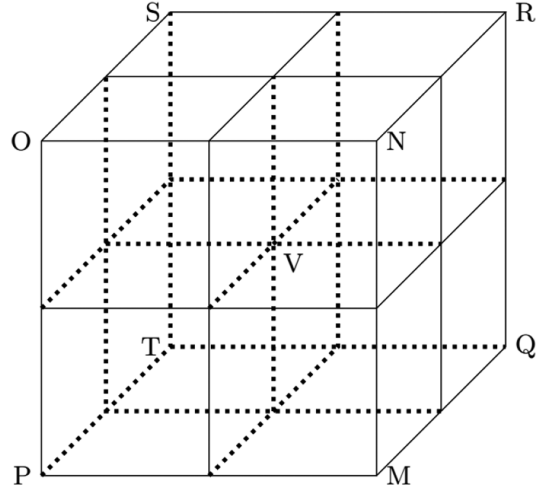

FIG. 4. $\Delta^* \phi_{ABCD} + \Delta^* \phi_{EFGH} + \Delta^* \phi_{IJKL}$ forms the closed surface of a big cube $OQ$, and every edge is canceled. On other types of lattices, the change of $S_{\text{CS}}$ under gauge transformation still has the form $-\sum_V \lambda_V \cdot$ closed surface, but the shape of the surface depends on the lattice.

The gauge invariance of the term

$$S_{\text{couple}} = \sum_{\text{adjacent cubes}} \bar{\psi}_{II} e^{i\bar{a}} \psi_I + \text{H.c.} \quad (16)$$

can be checked by substituting Eqs. (14) and (15) into Eq. (16).

When the gauge group is compact and the space has a nontrivial homotopy group, invariance under large gauge transformations will pose a constraint on the level coefficient $k$. For example, if the lattice is on a 3-torus with radius $R$, we do a large gauge transformation $a_\mu(x) \to a_\mu(x) + \Delta_\mu \lambda(x)$, where $\lambda(x + 2\pi R \hat{e}_0) = \lambda(x) + 2\pi$ and $\Delta_\mu \lambda(x) = \lambda(x + \hat{e}_\mu) - \lambda(x)$. Suppose the spatial surfaces enclose a unit flux. The change of the action (2) under such a large gauge transformation is

$$\delta S_{\text{CS}} = \frac{k}{2\pi} \sum_{x_0} \Delta_0 \lambda \sum_{x_1, x_2} \phi_0 = \frac{k}{2\pi} \times 2\pi \times 2\pi = 2k\pi, \quad (17)$$

where $\phi_0$ means a plaquette perpendicular to the 0 direction. The result tells us that $k$ has to be an integer in order to make $e^{iS_{\text{CS}}}$ gauge invariant. When the gauge group is noncompact, there is no large gauge transformation.

## IV. WILSON LOOPS

When matter fields are added to the action, the partition function is

$$Z = \int d\bar{\psi} d\psi da_\mu e^{i[S_{\text{CS}} + gS_{\text{couple}} + S_{\text{matter}}]}, \quad (18)$$

where $S_{\text{matter}}$ only contains matter fields, and $S_{\text{couple}}$ couples matter fields to the gauge field. When the coupling $g$ is small or $S_{\text{matter}}$ contains a large mass term, we can expand the partition function,

$$Z = \int d\bar{\psi} d\psi da_\mu \sum_n \frac{(ig)^n}{n!} S_{\text{couple}}^n e^{i[S_{\text{CS}} + S_{\text{matter}}]}. \quad (19)$$

If we substitute the definition of $S_{\text{couple}}$ [Eq. (16)] into the above Eq. (19), we find that the terms in $S_{\text{couple}}^n$ that do not form closed loops vanish. Both $S_{\text{CS}}$ and $da_\mu$ are invariant under the gauge transformation $a_\mu \to a_\mu + \Delta_\mu \lambda$, so only the terms in $S_{\text{couple}}^n$ that are also invariant under such a transformation will average to a nonzero value. This is Elitzur's theorem [20].

We separate the part that contains $a_\mu$ and define

$$Z_a[\bar{J}] = \int da e^{i[S_{\text{CS}} + \sum_{\text{all plaquettes}} \bar{J}^\mu \bar{a}_\mu]}, \quad (20)$$

where $\bar{J}^\mu$ penetrates a plaquette, and all $\bar{J}^\mu$ together form loops. $\bar{J}^\mu = n(\bar{J}^\mu = -n)$ if the loops penetrate the plaquette $n$ times along (opposite to) the plaquette direction. $\bar{J}^\mu = 0$ if the loops do not penetrate the plaquette. $\bar{a}_\mu$ also penetrates a plaquette. It is not a new field but the average of the eight vertical $a_\mu$ surrounding the plaquette. An example of the Wilson loop is shown in Fig. 5, and an example of $\bar{a}$ is shown in Fig. 1.

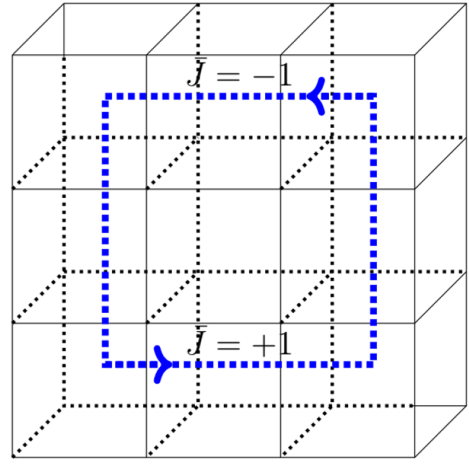

FIG. 5. The blue line on the dual lattice forms a Wilson loop, which is labeled by $\bar{J}$. $\bar{J}^\mu = n(\bar{J}^\mu = -n)$ if the Wilson loop penetrates the plaquette $n$ times along (opposite to) the plaquette direction. $\bar{J}^\mu = 0$ if the loop does not penetrate the plaquette.

The $\bar{a}$ field can also be labeled by the plaquette $p$ it penetrates, and the partition function (20) can be written as

$$Z_a[\bar{J}] = \int d\bar{a}_p e^{i \sum_p [\frac{k}{4\pi} \bar{a}_p \phi_p + \bar{J}_p \bar{a}_p]} \quad (21)$$

Note that $\bar{a}_p$ and $a_\mu$ are related by a nondegenerate linear transformation (3), so we can use either $\int da_\mu$ or $\int d\bar{a}_p$ to do the integration.

The saddle point equation for $\bar{a}_p$ is

$$\frac{k}{2\pi} \phi_p + \bar{J}_p = 0. \quad (22)$$

Substituting Eq. (22) into Eq. (21), we obtain the saddle point action,

$$\sum_p \left[ \frac{1}{2}(-\bar{J}_p)\bar{a}_p + \bar{J}_p \bar{a}_p \right] = \sum_p \frac{1}{2} \bar{J}_p \bar{a}_p. \quad (23)$$

Note that $\phi_p$ is a linear combination of $a_\mu$, so it is also a linear combination of $\bar{a}_p$. $\bar{a}_p \phi_p$ is actually quadratic with respect to $\bar{a}_p$, and the saddle point method should produce the exact result,

$$Z_a[\bar{J}] = Z_a[0] e^{i \sum_p \frac{1}{2} \bar{J}_p \bar{a}_p}, \quad (24)$$

where $\bar{a}_p$ is a linear combination of $a_p$ with coefficient $\frac{1}{8}$, so $\sum_p \bar{J}_p \bar{a}_p$ is just $\frac{1}{8}$ times a sum over loops on the original lattice. It can be represented by a sum over small plaquettes $q$ inside the loops on the original lattice,

$$Z_a[\bar{J}] = Z_a[0] e^{\frac{i}{16}\sum_q \phi_q} = Z_a[0] e^{-i\sum_q \frac{\pi}{8k} \bar{J}_q}. \quad (25)$$

The saddle point, Eq. (22), has been used in the second equality. This result counts how many times $\bar{J}$ penetrates the loops, and it reduces to the well-known expression in the continuous theory [2,21] when the lines do not intersect. When the lines intersect, however, fractional linking numbers will be allowed, but the result is still well defined, which is different from the continuous theory. Several examples are shown below:

(1) When two loops penetrate each other but do not intersect, the $a_\mu$ loops in $\sum_p \bar{J}_p \bar{a}_p$ are shown in Fig. 6. This figure can be obtained by replacing $\bar{a}_p$ in $\sum_p \bar{J}_p \bar{a}_p$ with $a_\mu$ using Eq. (3). On the dual lattice, the loops in $\sum_p \bar{J}_p \bar{a}_p$ consist of single lines, which is shown in Fig. 7. In Fig. 6, the loop of cubes is actually a weighted sum of line loops. Take the surface $ABCD$ for example, every edge is summed over for multiple times. In Fig. 8, the number of arrows represents the number of times each link $a_\mu$ is summed over. This sum can be reorganized into a sum over four loops, shown in Fig. 9.

The $\bar{J}$ loop $QRST$ in Fig. 7 goes through the center of $ABCD$ in Fig. 6, so it penetrates all of the four loops shown in Fig. 9. This contributes 4 to the sum $\sum_q \bar{J}_q$ in Eq. (25). In Fig. 6, surfaces $EFGH$, $IJKL$, and $MNOP$ are the same as surface $ABCD$, so the final result for the expectation value of the Wilson loop is

$$Z_a[\bar{J}] = Z_a[0] e^{-i\frac{\pi}{8k} 4 \times 4} = Z_a[0] e^{-i\frac{2\pi}{k}}. \quad (26)$$

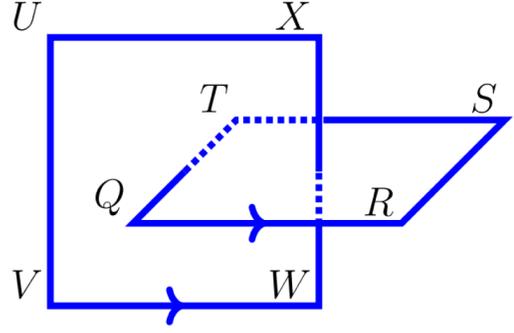

FIG. 7. The sum $\sum_p \bar{J}_p \bar{a}_p$ expressed on the dual lattice consists of two loops of lines. In this example, the two loops penetrate but do not intersect each other.

This agrees with the result in the continuous theory [2,21].

(2) When two loops of cubes intersect in a way shown in Fig. 10, the $\bar{J}$ loops going through the centers of the cubes are shown in Fig. 11. The loop decomposition of surface $ABCD$ in Fig. 10 is the same as Fig. 9, but this time the $\bar{J}$ loop $QRST$ shown in Fig. 11 only penetrates loop (i) and (iii) in Fig. 9. The same is true for surfaces $EFGH$, $IJKL$, and $MNOP$ in Fig. 10. So the final result for the expectation value of Wilson loop is

$$Z_a[\bar{J}] = Z_a[0] e^{-i\frac{\pi}{8k} 4 \times 2} = Z_a[0] e^{-i\frac{\pi}{k}}. \quad (27)$$

The linking number of the Wilson loops labeled by $\bar{J}_{QRST}, \bar{J}_{UVWX}$ in Fig. 11 is

$$\Phi[\bar{J}_{QRST}, \bar{J}_{UVWX}] = \frac{1}{2}. \qquad (28)$$

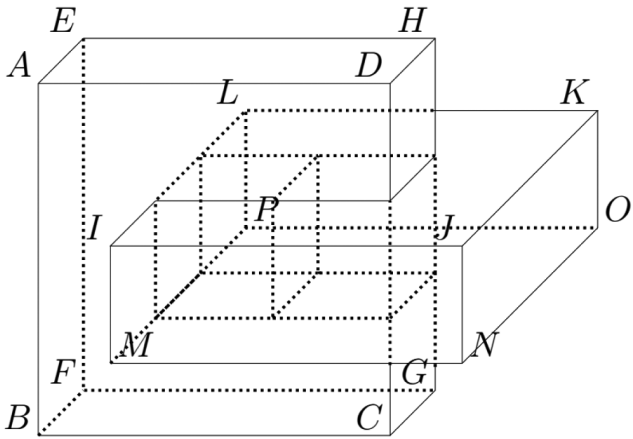

FIG. 6. When $\bar{J}$ labels two loops penetrating each other, the sum $\sum_p \bar{J}_p \bar{a}_p$ expressed on the original lattice consists of two loops of cubes.

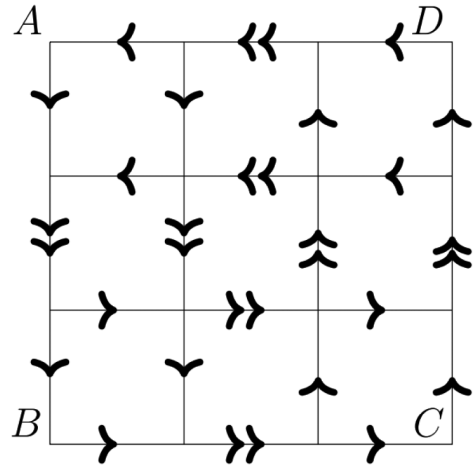

FIG. 8. The surface $ABCD$ in Fig. 6 is a sum of edges inside the surface. The number of arrows represents the number of times each edge is summed over.

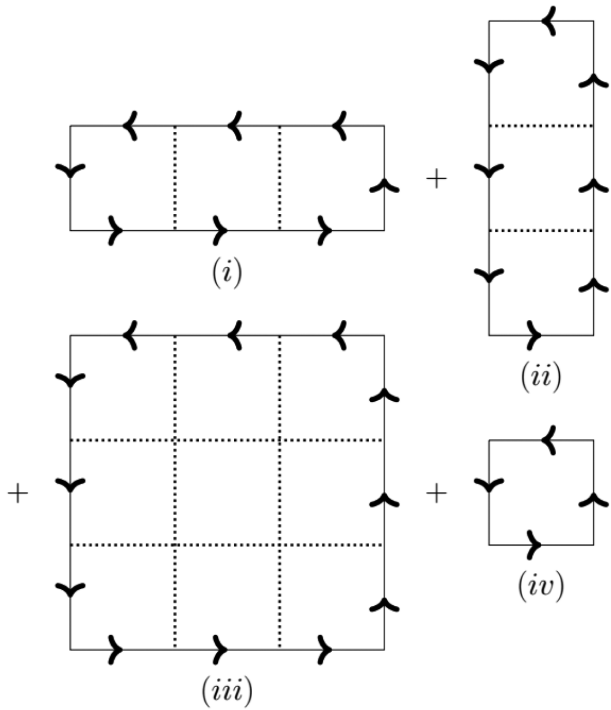

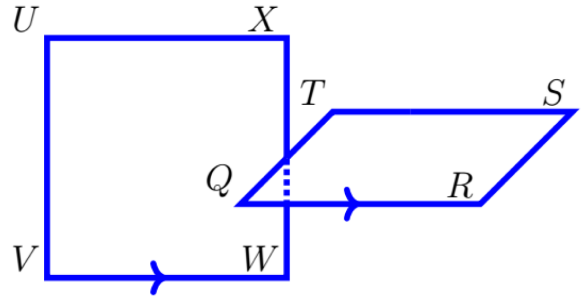

FIG. 11. The $\bar{J}$ loops of two Wilson loops interacting with each other. Since they are defined on a cubic lattice, the two loops are perpendicular to each other at the intersection point, and the linking number calculated from the lattice action (2) is $\frac{1}{2}$.

FIG. 9. The sum of edges in Fig. 8 is equivalent to a sum over four loops.

The appearance of the fractional linking number looks weird. The reader can check that whenever the Wilson loops intersect, the linking number calculated from the expectation value of Wilson loops is always a fraction, and it is not topological. As we change the angles between the $\bar{J}$ lines at the intersection point, fractions like $\frac{1}{4}, \frac{1}{8}$, etc. can appear. In Sec. VII, the lattice action is constructed on a triangular prism lattice and hexagonal prism lattice. Since there are angles $\frac{\pi}{3}, \frac{2\pi}{3}$ on these lattices, the fractional linking number can be $\frac{1}{3}, \frac{1}{6}, \frac{1}{12}$, etc. If we work on other types of lattices that allow different angles between the lines, we get even more fractions. The author believes that the nontopological result is a suggestion that the Chern-Simons theory in continuous spacetime is not well defined in these cases.

## V. THE CONTINUUM LIMIT

In order to compare the lattice construction with Chern-Simons theory in continuous spacetime, we restore the lattice spacing $\epsilon$ and let it approach 0. The error produced by replacing $a_p$, which is the gauge field living at the plaquette center, with $\bar{a}_p$ defined in Eq. (3), is of order $\epsilon^2$. The symmetry of the square plaquette cancels the first order error. If we further divide the lattice cubes' six surfaces into three pairs: front-back, left-right, up-down, and sum them separately, the Chern-Simons action defined in Eq. (2) reduces to

$$\begin{aligned}S_{CS} &= \frac{k\epsilon^3}{4\pi}\sum_{\text{cubes}}\left[\frac{1}{2}(\bar{a}_{\text{left}}\phi_{\text{left}}+\bar{a}_{\text{right}}\phi_{\text{right}})\right.\\ &\quad +\frac{1}{2}(\bar{a}_{\text{front}}\phi_{\text{front}}+\bar{a}_{\text{back}}\phi_{\text{back}})\\ &\quad \left.+\frac{1}{2}(\bar{a}_{\text{up}}\phi_{\text{up}}+\bar{a}_{\text{down}}\phi_{\text{down}})\right]\\ &= \frac{k\epsilon^3}{4\pi}\sum_{\text{cubes}}\frac{1}{2}[\vec{a}_{\text{left}}\cdot(\vec{\nabla}\times\vec{a}_{\text{left}})\\ &\quad +\vec{a}_{\text{right}}\cdot(\vec{\nabla}\times\vec{a}_{\text{right}})+\vec{a}_{\text{front}}\cdot(\vec{\nabla}\times\vec{a}_{\text{front}})\\ &\quad +\vec{a}_{\text{back}}\cdot(\vec{\nabla}\times\vec{a}_{\text{back}})+\vec{a}_{\text{up}}\cdot(\vec{\nabla}\times\vec{a}_{\text{up}})\\ &\quad +\vec{a}_{\text{down}}\cdot(\vec{\nabla}\times\vec{a}_{\text{down}})]+\mathcal{O}(\epsilon^5)\\ &= \frac{k\epsilon^3}{4\pi}\sum_{\text{cubes}}\vec{a}_{\text{center}}\cdot(\vec{\nabla}\times\vec{a}_{\text{center}})+\mathcal{O}(\epsilon^5)\\ &= \frac{k}{4\pi}\int d^3x\,\epsilon^{\mu\nu\rho}a_\mu\partial_\nu a_\rho+\mathcal{O}(\epsilon^5),\end{aligned}\qquad(29)$$

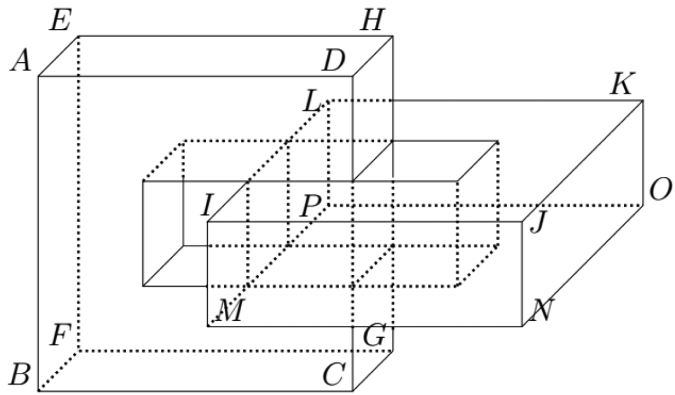

FIG. 10. When two Wilson loops intersect each other in a way shown in Fig. 11, the sum $\sum_p \bar{J}_p \bar{a}_p$ expressed on the original lattice are two loops of cubes intersecting each other.

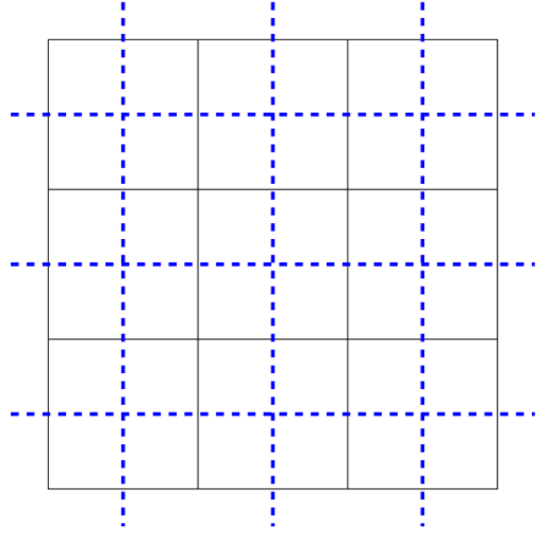

FIG. 12. If we take time to be continuous, the cubic lattice becomes the square lattice. The blue dashed line is the dual lattice.

where $\vec{a}_{\text{left}}$ means the three-vector $\vec{a}$ at the center of the left surface. The error at order $\epsilon^4$ is again canceled by symmetry, and the lattice model agrees with the continuous theory up to order $\epsilon^5$. One can also construct the lattice Chern-Simons theory on a lattice formed by simplexes and use a cup product to stick together the 1-cochain $a$ and 2-cochain $\delta a$. These are the standard language of algebraic geometry, and the gauge invariance is guaranteed. One such example is shown in [15]. However, the difference between this model and the continuous theory is of order $\epsilon^4$. Actually, if we only require the lattice model to agree with the continuous theory up to order $\epsilon^4$ and be gauge invariant at the leading order, the easiest way is to replace $\bar{a}_p$ in Eq. (2) with any one of the vertical links surrounding the plaquette in Fig. 1.

## VI. HAMILTONIAN FORMALISM

If we only take time to be continuous, the cubic lattice becomes a square lattice (Fig. 12). Since the cube is compressed into a square, the relation between $\bar{a}$ and $a$ also changes. In Fig. 13, Eq. (3) becomes

$$\bar{a}_{GH}=\frac{1}{4}(a_{AB}+a_{BC}+a_{DE}+a_{EF}),$$
$$\bar{a}_0(H)=\frac{1}{4}[a_0(B)+a_0(C)+a_0(F)+a_0(E)].\qquad(30)$$

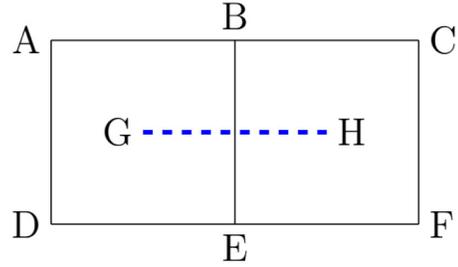

FIG. 13. Since the cube is compressed into a square, the relation between $\bar{a}$ and $a$ becomes $\bar{a}_{GH}=\frac{1}{4}(a_{AB}+a_{BC}+a_{DE}+a_{EF})$, $\bar{a}_0(H)=\frac{1}{4}[a_0(B)+a_0(C)+a_0(F)+a_0(E)]$.

Equation (2) becomes

$$\begin{aligned}S_{CS}&=\frac{k}{4\pi}\sum_{x_1,x_2}\int dx_0\bar{a}_0\phi_0+\bar{a}_1(\Delta_2 a_0-\dot{a}_2)\\ &\quad +\bar{a}_2(\dot{a}_1-\Delta_1 a_0)\\ &=\frac{k}{4\pi}\sum_{x_1,x_2}\int dx_0 a_0\bar{\phi}_0+a_0\Delta_1\bar{a}_2-a_0\Delta_2\overline{a_1}-\bar{a}_1\dot{a}_2\\ &\quad +\bar{a}_2\dot{a}_1\\ &=\frac{k}{4\pi}\sum_{x_1,x_2}\int dx_0 2a_0\bar{\phi}_0-\bar{a}_1\dot{a}_2+\bar{a}_2\dot{a}_1\\ &=\frac{k}{2\pi}\sum_{x_1,x_2}\int dx_0 a_0\bar{\phi}_0-\bar{a}_1\dot{a}_2\\ &=\frac{k}{2\pi}\sum_{x_1,x_2}\int dx_0 a_0\bar{\phi}_0+\bar{a}_2\dot{a}_1,\end{aligned}\qquad(31)$$

where we have put in the explicit expression for $\phi$,

$$\phi_\mu = \epsilon^{\mu\nu\rho}\Delta_\nu a_\rho, \tag{32}$$

and used the duality between the lattice and dual lattice to write $\sum_{\vec{x}} \bar{a}_0 \phi_0 = \sum_{\vec{x}} a_0 \bar{\phi}_0$, $\sum_{\vec{x}} \bar{a}_1 \dot{\bar{a}}_2 = \sum_{\vec{x}} \bar{a}_1 \dot{a}_2$, which can also be easily checked by substituting $\bar{a}, \bar{\phi}_0$ with their explicit expressions [Eq. (30)]. Note that $a_0$ is purely a Lagrangian multiplier. Integrating over $a_0$ generates the constraint $\bar{\phi}_0 = 0$ for every spatial plaquette, which also means $\phi_0 = 0$. The Chern-Simons Lagrangian becomes

$$\mathcal{L}_{CS} = \frac{k}{2\pi}\bar{a}_2\dot{a}_1 \quad \text{or} \quad -\frac{k}{2\pi}\bar{a}_1\dot{a}_2 \quad \text{with}$$
$$\bar{\phi}_0 = \phi_0 = 0. \tag{33}$$

Note that the Lagrangian is linear in time derivative, so $\frac{k}{2\pi}\bar{a}_2$ is the conjugate variable of $a_1$ and $-\frac{k}{2\pi}\bar{a}_1$ is the conjugate variable of $a_2$. The Hamiltonian vanishes,

$$H_{CS} = 0, \tag{34}$$

so every state has zero energy. Upon quantization, the commutation relation is

$$[a_1(\vec{x}), \bar{a}_2(\vec{y})] = \frac{2\pi i}{k}\delta_{\vec{x},\vec{y}}, \quad [a_2(\vec{x}), \bar{a}_1(\vec{y})] = -\frac{2\pi i}{k}\delta_{\vec{x},\vec{y}}. \tag{35}$$

But this does not means that $a_1$ and $a_2$ commute, because $\bar{a}_1$ and $\bar{a}_2$ are linear combinations of $a_1, a_2$. In Fig. 13, the commutation relation is

$$[a_{EB}, \bar{a}_{GH}] = -\frac{2\pi i}{k}$$
$$\Leftrightarrow [a_{EB}, a_{AB}] = [a_{EB}, a_{BC}]$$
$$= [a_{EB}, a_{DE}] = [a_{EB}, a_{EF}] = -\frac{\pi i}{2k}. \tag{36}$$

If the space has a trivial homotopy, the constraint $\phi_0 = 0$ uniquely determines the physical state, so there is only one state in the Hilbert space. When the space has a nontrivial homotopy group, the states are degenerate. For example, if the space is a torus (Fig. 14), the gauge-invariant operators that classify the physical states are the two orthogonal Wilson loops,

$$W_1 = e^{i\sum_{x_1} a_1}, \quad W_2 = e^{i\sum_{x_2} a_2}. \tag{37}$$

One can also construct Wilson loop operators on the dual lattice, $\bar{W}_1 = e^{i\sum_{x_1} \bar{a}_1}$, for example. However, they are not independent because they are actually products of the $W$ operators.

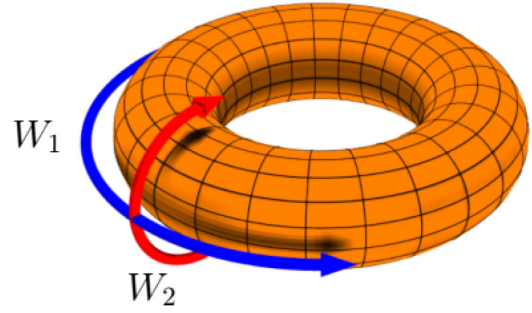

FIG. 14. The gauge-invariant operators on a torus are the two orthogonal Wilson loops.

$W_1, W_2$ do not commute at the intersection point. Using the Baker-Campbell-Hausdorff formula, $e^X e^Y = e^{X+Y+\frac{1}{2}[X,Y]}$, and in the commutation relation [Eq. (36)], we find that

$$W_1 W_2 = e^{-\frac{2\pi i}{k}} W_2 W_1. \tag{38}$$

If we find an eigenstate $|s\rangle$ for $W_1$, the other eigenstates can be generated by $W_2^n|s\rangle$ where $0 < n < k$, so the Hilbert space is k dimensional. Physically, acting $W_2$ on $|s\rangle$ means inserting a flux quanta vertically through the center of the torus. This does not violate the constraint $\phi_0 = 0$ because the flux does not penetrate the surface. In general, the dimension of the Hilbert space for a genus-g surface is $k^g$, which agrees with the continuous theory. The states do not duplicate, because although we have two lattices, we really have one set of variables. We either use $a$ or $\bar{a}$, one can be expressed in terms of the other.

## VII. GENERALIZATIONS

The key expression of this paper is

$$S_{CS} = \frac{k}{4\pi}\sum_p \bar{a}_p \phi_p. \tag{39}$$

This applies to many different lattices, and the explicit form of $\bar{a}_p$ depends on the lattice type. It is straightforward to generalize this construction to other polygonal prism lattices. In this section is shown the constructions of lattice Chern-Simons theory on a triangular prism lattice and hexagonal prism lattice. All of the lattice links are assumed to have a unit length for simplicity.

There are two kinds of plaquettes on the triangular prism lattice: equilateral triangle and square. When the plaquette $p$ in $\bar{a}_p \phi_p$ is the triangle, as is shown in Fig. 15, then obviously

$$\bar{a}_{DEF} = \frac{1}{6}(a_{GD} + a_{DA} + a_{HE} + a_{EB} + a_{IF} + a_{FC}) \tag{40}$$

$$\phi_{DEF} = a_{DE} + a_{EF} + a_{FD}. \tag{41}$$

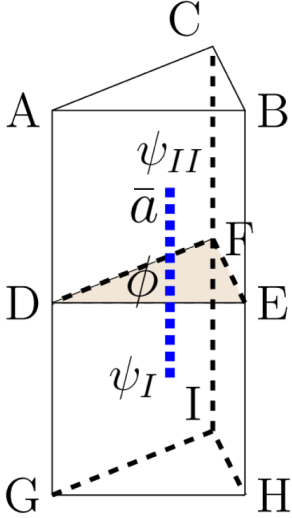

FIG. 15. When the plaquette $p$ is an equilaterial triangle, the $\bar{a}_p$ penetrating the plaquette is the average of the six vertical links surrounding the plaquette. Matter fields couple to gauge fields through the dual link.

When the plaquette is a square, as is shown in Fig. 16, the expression for $\bar{a}$ needs a little more consideration. First, the length of the dual link $\bar{a}$, which connects the centers of two adjacent prisms, is not 1, but $\frac{1}{\sqrt{3}}$ (assume the length of every link on the original lattice is 1). Second, the angle between $\bar{a}$ and the eight links $AB, AD, EF, EH, FG, HG, BC, DC$ is not 0 but $\pi/6$. The average of the eight links actually calculates $\cos(\pi/6)\frac{\bar{a}}{1/\sqrt{3}}$, because the components of the eight links perpendicular to the dual link are canceled upon summation, and the length of the dual link has to be rescaled to be the same as the eight links. Inverse the above analysis, we have

$$\bar{a}_{BFHD} = \frac{1}{\sqrt{3}}\frac{1}{\cos(\pi/6)}\frac{1}{8}(a_{AB} + a_{AD} + a_{EF} + a_{EH} + a_{BC}$$
$$+ a_{DC} + a_{FG} + a_{HG})$$
$$= \frac{1}{12}(a_{AB} + a_{AD} + a_{EF} + a_{EH} + a_{BC} + a_{DC}$$
$$+ a_{FG} + a_{HG}), \quad (42)$$

$$\phi_{BFHD} = a_{BF} + a_{FH} + a_{HD} + a_{DB}. \quad (43)$$

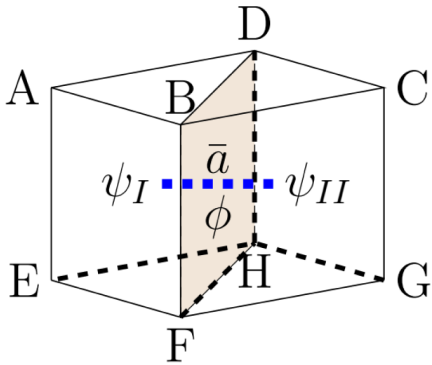

FIG. 16. When the plaquette $p$ is a square, $\bar{a}_p$ is not a direct average of the eight links surrounding the plaquette. The length of the dual link and the angle between $\bar{a}$ and the surrounding links have to be taken into consideration. Matter fields couple to gauge fields through the dual link.

Matter fields live on the dual lattice, and they are connected via the dual links. As is shown in Figs. 15 and 16, matter fields couple to gauge fields through the term

$$S_{\text{couple}} = \sum_{\text{adjacent cells}} \psi_I e^{i\bar{a}} \bar{\psi}_{II} + \text{H.c.} \quad (44)$$

The analog of Fig. 4 on this lattice is Fig. 17. Gauge invariance is guaranteed, because the change of the action under gauge transformation is

$$\delta S_{\text{CS}} = -\frac{k}{4\pi}\sum_V \frac{\lambda_V}{6}(\phi_{ABHG} + \phi_{BCIH} + \phi_{CDJI} + \phi_{DEKJ}$$
$$+ \phi_{EFLK} + \phi_{FAGL} + \phi_{AFEDCB} + \phi_{GHIJKL})$$
$$= 0. \quad (45)$$

The same analysis also applies to the hexagonal prism lattice. There are also two kinds of plaquettes: hexagon and square. If the plaquette that $\bar{a}$ penetrates is a hexagon shown in Fig. 18, the links surrounding the plaquette are perpendicular to the plaquette, and $\bar{a}$ has unit length. So $\bar{a}_p$

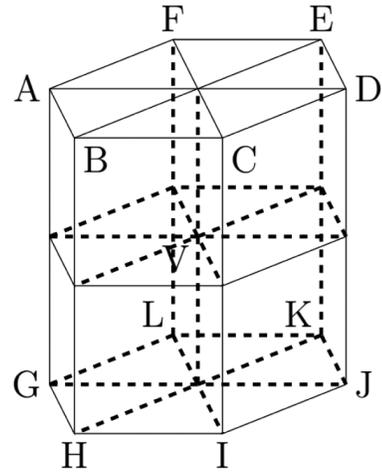

FIG. 17. The change of $S_{\text{CS}}$ under gauge transformation still has the form $-\sum_V \lambda_V \cdot$ closed surface. Every edge is canceled on the closed surface so the action is exactly gauge invariant.

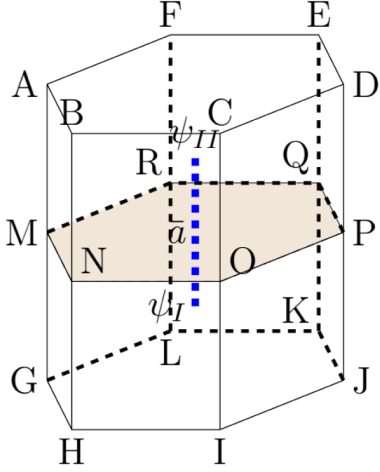

FIG. 18. When the plaquette $p$ is a hexagon, $\bar{a}_p$ penetrating the plaquette is the average of the 12 links surrounding the plaquette.

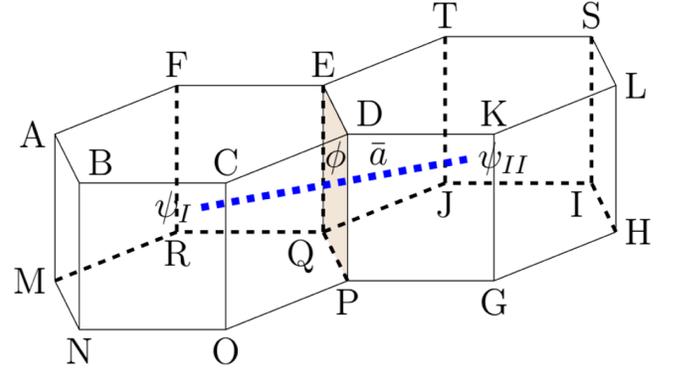

FIG. 19. When the plaquette is a square on the hexagonal prism lattice, the angle between the dual link and the surrounding links is $\frac{\pi}{6}$, and the length of the dual link is $\sqrt{3}$. These have to be taken into account when we write down $\bar{a}_p$.

is the direct average of the 12 links surrounding the plaquette,

$$\bar{a}_{MNOPQR} = \frac{1}{12}(a_{MA} + a_{NB} + a_{OC} + a_{PD} + a_{QE} \\ + a_{RF} + a_{GM} + a_{HN} + a_{IO} + a_{JP} \\ + a_{KQ} + a_{LR}), \quad (46)$$

$$\phi_{MNOPQR} = a_{MN} + a_{NO} + a_{OP} + a_{PQ} + a_{QR} + a_{RM}. \quad (47)$$

If the plaquette is a square shown in Fig. 19, the eight links surrounding the plaquette are not perpendicular to the plaquette. The angle between these links and the dual link is $\pi/6$. The length of the dual link connecting centers of the two prisms is $\sqrt{3}$. So we have

$$\bar{a}_{DEQP} = \sqrt{3}\frac{1}{\cos(\pi/6)}\frac{1}{8}(a_{EF} + a_{DC} + a_{PO} + a_{QR} + a_{GP} \\ + a_{JQ} + a_{TE} + a_{KD}) \\ = \frac{1}{4}(a_{EF} + a_{DC} + a_{PO} + a_{QR} + a_{GP} + a_{JQ} + a_{TE} \\ + a_{KD}), \quad (48)$$

$$\phi_{DEQP} = a_{DE} + a_{EQ} + a_{QP} + a_{PD}. \quad (49)$$

The form of the coupling term $S_{\text{couple}}$ is the same as Eq. (5). Generalizations to other lattices with multiple types of polygonal prisms follow exactly the same method shown in this section.

## VIII. SUMMARY AND OUTLOOK

In this paper, a lattice version of Abelian Chern-Simons theory is defined. In order to maintain the Leibniz rule on the lattice and preserve gauge invariance, this proposal introduces the dual lattice and expresses gauge fields on the dual lattice in terms of the gauge fields on the original lattice. Upon gauge transformation, the matter fields on the dual lattice transform as a collection of parton fields on the vertexes of the lattice cell. These treatments circumvent the issue of forward/backward difference and also avoids the duplication problem. The construction agrees with the continuous theory up to order $\mathcal{O}(\epsilon^5)$, which is higher than the existing constructions. The form of the action is symmetric with respect to the three spacetime dimensions, and the calculation of Wilson loops is straightforward. The Wilson loops are loops of cubes on the lattice, and they can be decomposed into several line loops. This is similar to the idea of treating Wilson loops as ribbons to regularize the framing anomaly [3,22]. When the Wilson loops do not intersect, the expectation value is the same as the Chern-Simons theory in continuous spacetime and is related to the linking number. Generalizations to triangular prism and hexagonal prism lattices are also discussed.

Finally, there are still open questions to be answered. One problem is to construct lattice Chern-Simons theory on any irregular three-dimensional spacetime lattices and to include non-Abelian gauge groups. Whether there is a topological-invariant way to define the value of crossing Wilson loops on the lattice or their fractional expectation values have a unique expression in terms of the angles at the intersection point are still waiting to be discussed.

## ACKNOWLEDGMENTS

The author would like to thank Prof. Tom Banks for helpful discussions. This work is supported in part by the DOE under Grant No. DE-SC0010008.